\documentclass[pre,preprintnumbers,amsmath,amssymb]{revtex4}
\usepackage{graphicx}
\usepackage{bm}

\begin{document}
\title{Maximum Distance Between the Leader and the Laggard for Three 
Brownian Walkers}

\author{Satya N. Majumdar}
\affiliation{Univ. Paris Sud, CNRS, LPTMS,
UMR 8626, Orsay F-91405, France}

\author{Alan J. Bray}
\affiliation{School of Physics and Astronomy, University of Manchester, 
Manchester M13 9PL,UK}

\begin{abstract}
We consider three independent Brownian walkers moving on a line. The process 
terminates when the left-most walker (the `Leader') meets either of the 
other two walkers. For arbitrary values of the diffusion 
constants $D_1$ (the Leader), $D_2$ and $D_3$ of the three walkers, 
we compute the probability distribution $P(m|y_2,y_3)$ of the maximum 
distance $m$ between the Leader and the current right-most particle 
(the `Laggard') during the process, where $y_2$ and $y_3$ are the initial  
distances between the leader and the other two walkers. 
The result has, for large $m$, the form 
$P(m|y_2,y_3) \sim A(y_2,y_3)\,m^{-\delta}$, where 
$\delta = (2\pi-\theta)/(\pi-\theta)$ and 
$\theta = \cos^{-1}(D_1/\sqrt{(D_1+D_2)(D_1+D_3)}$. The amplitude 
$A(y_2,y_3)$ is also determined exactly.  
\end{abstract}

\maketitle
\date{\today}

\section{Introduction}
The  unions of  reactions of  three diffusing  particles  (i.e.\ three
Brownian   walkers)  have   been  much   studied  in   the  literature
\cite{RednerBook}. Such systems are  often amenable to exact solution,
even  for arbitrary values  of the  diffusion constants  $D_1$, $D_2$,
$D_3$,  of  the  particles,  whereas  systems  with  more  than  three
particles are not analytically tractable (one exception being the case
when the particles are mutually annihilating, i.e.\ `vicious walkers',
with equal diffusion constants  \cite{FisherHuse}).  An example of the
type  of three-particle  problem that  can  be exactly  solved is  the
computation  of  the probability  that  the  left-most particle  (with
diffusion constant $D_1$) has not  been touched by either of the other
two particles up to time $t$.  This probability has a power-law decay,
$P \sim t^{-\theta_1}$, with 
\cite{RednerBook,benAvraham,FisherGelfand,benAvraham2}
\begin{equation}
\theta_1 = \frac{\pi}{2(\pi-\theta)}
\end{equation}
where 
\begin{equation}
\theta = \cos^{-1}\left[\frac{D_1}{\sqrt{(D_1+D_2)(D_1+D_3)}}\right].
\label{theta}
\end{equation}
In  this paper  we consider  a  related aspect  of the  three-particle
system that  has not  been addressed so far. 
We define  the initially
left-most  of  the  three  particles  to  be  the  `Leader',  and  the
right-most of  the remaining  two particles to  be the  `Laggard' (the
Leader-Laggard  terminology  was   introduced  by  ben Avraham  et  al.\ 
\cite{benAvraham2}), and we again consider processes which terminate when 
the Leader is touched by either of the other two particles. We compute 
the probability  distribution,  over this  set  of  processes (with  given
initial  conditions  for  the   particle  locations)  of  the  maximum
distance, $m$, between  the Leader and the Laggard.  We find that this
probability   distribution  has   a   power-law  tail   of  the   form
$m^{-\delta}$, where  $\delta$ is a nontrivial function  of the walker
diffusion constants $D_1$, $D_2$ and $D_3$. We note that for the special
case $D_1=0$, this distribution of the maximal distance between the Leader
and the Laggard was recently computed for arbitrary $N\ge 1$ independent 
particles~\cite{KMR}. However, for $D_1>0$, it is not easy to
generalise this method for arbitrary $N$ and in this paper we show
that the exact solution even for the $N=3$ case is highly nontrivial. 

\begin{figure}
\includegraphics[height=8.0cm,width=8.0cm,angle=0]{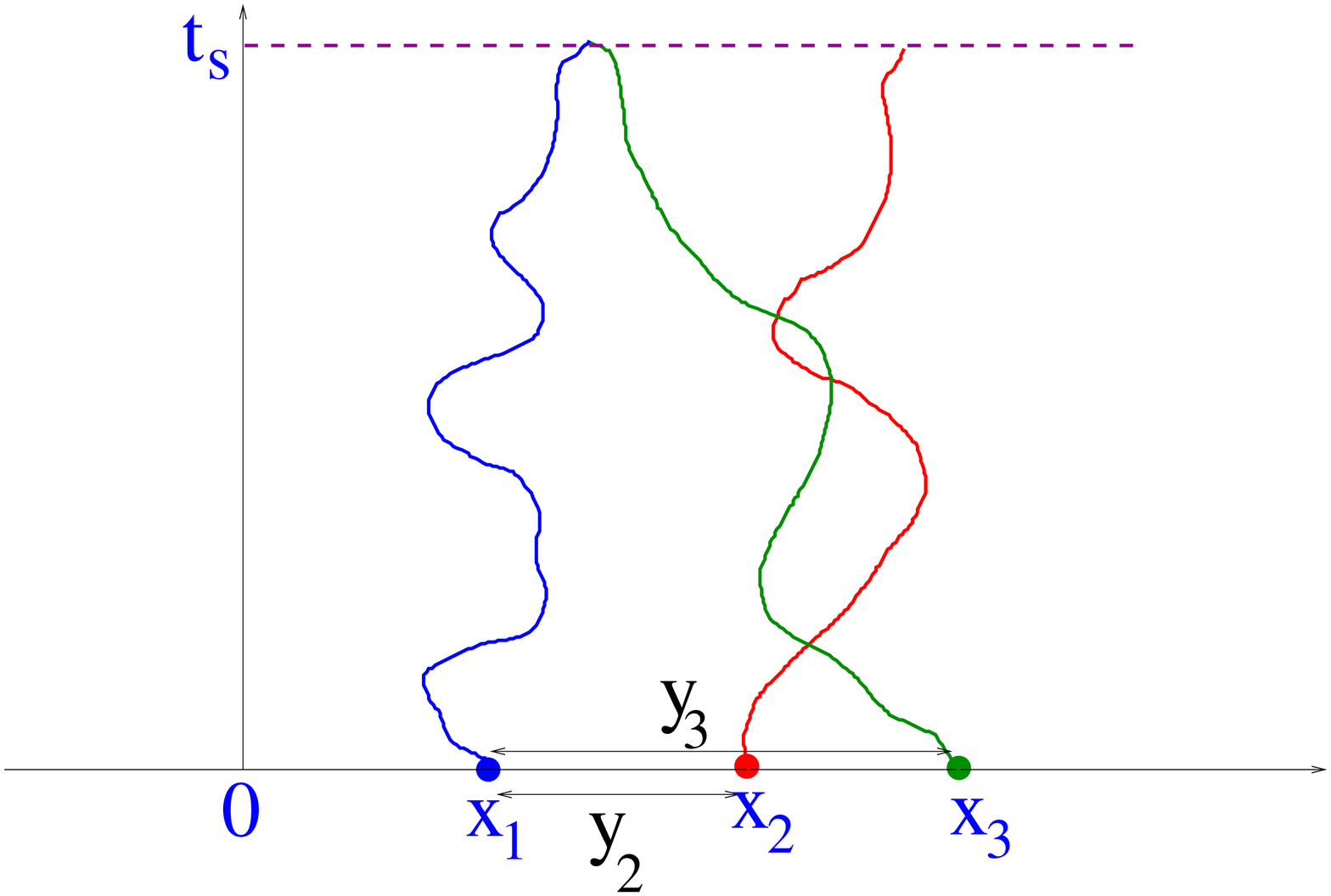}
\caption{\label{fig:n3} The trajectories of $3$ independent
Brownian walkers with initial positions $x_1$, $x_2$ and $x_3$.
The process stops at the stopping time $t_s$ when the
leftmost (blue) particle meets any other particle, such as
the third (green) particle in the figure.}
\end{figure}

Thus we consider three Brownian particles on a line with positions 
$\{x_1(t),x_2(t),x_3(t)\}$ that evolve independently with time according 
to the Langevin equations
\begin{equation}
\frac{dx_i}{dt}= \eta_i(t)
\label{evol1}
\end{equation}
where  $\eta_i(t)$ ($i=1$, $2$ or $3$) are independent 
Gaussian white noises with zero mean $\langle \eta_i(t)\rangle=0$ 
and the two-time correlator, $\langle \eta_i(t)\eta_j(t')\rangle = 2D_i 
\delta_{i,j} \delta(t-t')$. Thus $D_i$ denotes the diffusion constant
of the $i$-th particle. Let the initial positions of the three particles
be denoted by $x_1(0)=x_1$, $x_2(0)=x_2$ and $x_3(0)=x_3$ where
$x_1\le x_2\le x_3$ (see Fig. \ref{fig:n3}). 

Let $y_2(t)= x_2(t)-x_1(t)$ denote 
the separation at time $t$ between the first and the second
particle. Similarly $y_3(t)=x_3(t)-x_1(t)$ denotes the separation
at time $t$ between the first and the third particle. These
relative coordinates start respectively from their initial values 
$y_2(0)=y_2=x_2-x_1\ge 0$ and $y_3(0)=y_3= x_3-x_1$ (see Fig. \ref{fig:n3}), 
and subsequently evolve in time via
\begin{eqnarray}
\frac{dy_2}{dt}&=& \eta_2(t)-\eta_1(t)= \xi_2(t) \label{evol2} \\
\frac{dy_3}{dt}&=& \eta_3(t)-\eta_1(t)= \xi_3(t) \label{evol3} 
\end{eqnarray}
where the two noises $\xi_2(t)$ and $\xi_3(t)$ are now correlated
for $D_1>0$. Clearly $\langle \xi_2(t)\rangle= \langle \xi_3(t)\rangle =0$, 
while the two-time correlators are given by
\begin{eqnarray}
\langle \xi_2(t)\xi_2(t')\rangle &=& 2(D_1+D_2)\delta(t-t'), \label{noise2}\\
\langle \xi_3(t)\xi_3(t')\rangle &=& 2(D_1+D_3) \delta(t-t'), \label{noise3}\\
\langle \xi_2(t)\xi_3(t')\rangle &=& 2D_1 \delta(t-t'). \label{noise23}
\end{eqnarray}

Let $z(t)= {\rm max}\left(y_2(t), y_3(t)\right)$ denote the span
of this $3$-particle process at time $t$, i.e., $z(t)$ denotes the distance at 
time $t$ between the left-most (the Leader) and the right-most (the Laggard)
particles. We stop the process at a stopping time $t_s$ when the 
leader meets, for the first time, any of the other two particles (see
Fig. \ref{fig:n3} for an example of a realization of the process). 

Let 
\begin{equation}
m= \max_{0\le t\le t_s}\left[z(t)\right]
\label{max1}
\end{equation}
denote the maximum
value of the span till the stopping time $t_s$. 
Note that both $t_s$ and $m$ change from one realization of the
process to another. These two random variables are clearly
correlated. Here we are interested in the probability distribution 
(marginal) of $m$ only, i.e.\ , $P(m|y_2,y_3)$ given the initial 
separations $y_2$ and $y_3$.
We will show that $P(m|y_2,y_3)$ has a power law tail for large $m$
\begin{equation}
P(m|y_2,y_3) \simeq \frac{A(y_2,y_3)}{m^{\delta}}
\label{pdf1}
\end{equation}
where the exponent $\delta$ depends continuously on the
three diffusion constants $D_1$, $D_2$ and $D_3$ and has
the following exact expression
\begin{equation}
\delta= \frac{2\pi-\theta}{\pi-\theta}\ ,
\label{expon1}
\end{equation}
where $\theta$ is given by Eq.\ (\ref{theta})

We also compute the amplitude $A(y_2,y_3)$ of this power law decay
exactly. This amplitude is evidently a symmetric function of $y_2$
and $y_3$ but its explicit expression turns out to be rather nontrivial.

\section{Derivation of the Result}
To derive our result, it turns out to be more convenient to consider
the cumulative distribution of the maximum $m$ denoted by
\begin{equation}
F(y_2,y_3|L)= \int_0^L P(m|y_2,y_3)\, dm.
\label{cum1}
\end{equation}
Thus $F(y_2,y_3|L)$ denotes the probability that the maximum span
does not exceed $L$ till the stopping time $t_s$, given the initial
separations $y_2$ and $y_3$. The idea is to
write down a backward differential equation (Backward Fokker-Planck 
equation) for $F(y_2,y_3|L)$, treating the initial separations $y_2$ and 
$y_3$ as the independent variables.
To do this, we consider a typical evolution of the joint process
$\{y_2(t),y_3(t)\}$ via the Langevin equations (\ref{evol2})
and (\ref{evol3}), starting from the initial values $\{y_2,y_3\}$. 
Let us split the full time interval $[0,t_s]$ of the evolution into
two parts: over an initial infinitesimal time window $[0,\Delta t]$
where the joint process $\{y_2(t),y_3(t)\}$ evolves from
its initial value $\{y_2,y_3\}$ to the new value $\{y_2+\Delta y_2, 
y_3 + \Delta y_3\}$, where 
\begin{equation}
\Delta y_i = \int_0^{\Delta t} \xi_i(t) dt\ \ \ \ (i=2,3)\ ,
\end{equation}
and a subsequent interval $[\Delta t, t_s]$ where
the process evolves starting from its `new' initial position
$\{y_2 + \Delta y_2, y_3 + \Delta y_3\}$. 
Using the Markov property of the evolution, it then follows that
\begin{equation}
F(y_2,y_3|L)= \langle F\left(y_2 + \Delta y_2, y_3 +\Delta y_3|L\right) \rangle
\label{bfp0}
\end{equation}
where the angled brackets denotes the average over the initial displacements 
$\Delta y_2$ and $\Delta y_3$. 

We next expand the right-hand side of Eq. (\ref{bfp0}) in a Taylor series in 
$\Delta t$ (to first order in $\Delta t)$ using (i) $\langle 
\Delta y_i \rangle =0$ (for $i=2,3$) and (ii) the following covariances 
(which follow from the delta correlators in Eqs. (\ref{noise2}), 
(\ref{noise3}) and  (\ref{noise23}))
\begin{eqnarray}
\langle (\Delta y_2)^2 \rangle & = & 2\,(D_1+D_2)\Delta t  \\
\langle (\Delta y_3)^2 \rangle & = & 2\,(D_1+D_3)\Delta t  \\
\langle (\Delta y_2\,\Delta y_3)\rangle & = & 2D_1\,\Delta t .
\label{covar1}
\end{eqnarray}
Keeping only terms of  $O(\Delta t)$ then gives us the following partial 
differential equation for $F(y_2,y_3|L)$: 
\begin{equation}
(D_1+D_2)\frac{\partial^2F}{\partial y_2^2} 
+(D_1+D_3)\frac{\partial^2F}{\partial y_3^2}
+2D_1 \frac{\partial^2F}{\partial y_2 \partial y_3}=0.
\label{bfp}
\end{equation}

Note that the information that the process stops at a certain stopping time 
$t_s$ is actually captured only through the boundary conditions.
Eq. (\ref{bfp}) holds over the square $0\le y_2 \le L$ and $0\le y_3\le L$
in the two dimensional $(y_2,y_3)$ plane with the following boundary conditions
\begin{eqnarray}
F(y_2=0,y_3|L)&=& 1 \label{bc1} \\
F(y_2, y_3=0|L) &=& 1 \label{bc2} \\
F(y_2=L, y_3|L) &=& 0 \label{bc3}\\
F(y_2, y_3=L|L) &= & 0. \label{bc4}
\end{eqnarray} 
For example, if the initial separation $y_2=0$ and $0\le y_3\le L$, then the 
process stops
immediately, i.e., $t_s=0$, since the second particle has already hit
the leftmost particle. Clearly 
then the maximum $m=y_3$ which, with probability $1$, is less than or equal to 
$L$. Hence the boundary condition \eqref{bc1}. By symmetry, \eqref{bc2} 
follows. In contrast, if initially say $y_2=L$ and $0\le y_3\le L$, clearly
the initial value of $m$ is already $L$. So, the probability that $m$ will
stay below $L$ subsequently is clearly $0$, indicating the boundary condition
\eqref{bc3}. By symmetry, one then has \eqref{bc4}.
So, the technical challenge is now to solve the partial differential
equation \eqref{bfp} inside the square $[0,L]\times [0,L]$ with the above 
boundary conditions in Eqs. (\ref{bc1})-(\ref{bc4}).

To proceed, we make a linear transformation that gets rid of the cross term
in Eq. (\ref{bfp}). In other words, we diagonalize the covariance matrix. 
It turns out that a linear transformation that does the job is given by
\begin{eqnarray}
W_2 &=& \frac{1}{\sqrt{2+\gamma}}\left(\frac{y_2}{\sqrt{D_1+D_2}}+ 
\frac{y_3}{\sqrt{D_1+D_3}}\right) \label{w2} \\
W_3 &=& \frac{1}{\sqrt{2-\gamma}}\left(-\frac{y_2}{\sqrt{D_1+D_2}}+
\frac{y_3}{\sqrt{D_1+D_3}}\right) \label{w3} 
\end{eqnarray}
where
\begin{equation}
\gamma= \frac{2D_1}{\sqrt{(D_1+D_2)(D_1+D_3)}}.
\label{gamma}
\end{equation}
It is easy to see that $0\le \gamma\le 2$ for all $D_1\ge 0$, $D_2\ge 0$ and 
$D_3\ge 0$.
Note also that exchanging $y_2$ and $y_3$ and also $D_2$ and $D_3$,
is equivalent to letting $W_2\to W_2$ and $W_3\to -W_3$.  

In terms of these new variables $(W_2,W_3)$, Eq. (\ref{bfp}) becomes 
Laplace's equation 
\begin{equation}
\frac{\partial^2F}{\partial W_2^2}
+\frac{\partial^2F}{\partial W_3^2}=0.
\label{bfp1}
\end{equation}
\begin{figure}
\includegraphics[height=12.0cm,width=12.0cm,angle=0]{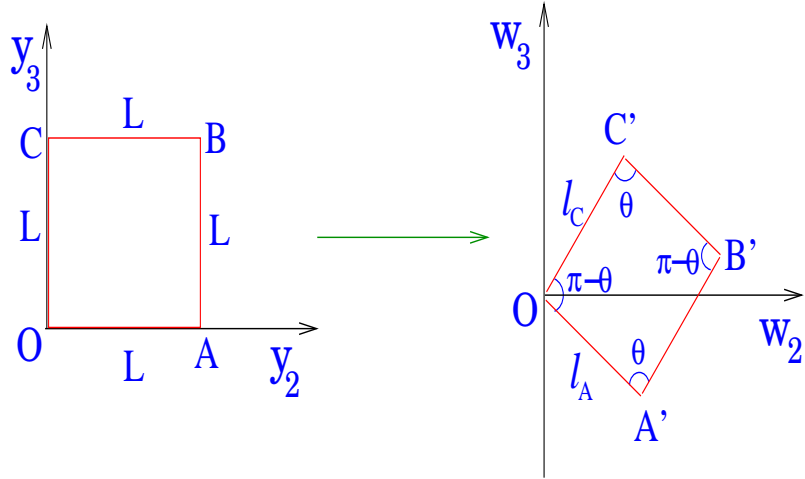}
\caption{\label{fig:lt} The linear transformation in Eqs. (\ref{w2})
and (\ref{w3}) from $(y_2,y_3)$ to the $(W_2,W_3)$ plane consists
of rotation and stretching. The square $[0,L]\times [0,L]$ with
vertices $0$, $A$, $B$ and $C$ in the
$(y_2,y_3)$ plane transforms to a parallelogram 
with vertices $O$, $A'$, $B'$ and $C'$ in the $(W_2,W_3)$ plane.}
\end{figure}

The original square $[0,L]\times [0,L]$ in the $(y_2,y_3)$ plane
transforms into a parallelogram in the $(W_2,W_3)$ plane under
the linear transformation in Eqs. (\ref{w2})
and (\ref{w3}) (see Fig. \ref{fig:lt}). 
The vertices $(O,A,B,C) \to (O,A',B',C')$ under this transformation.
It is easy to check that the lengths of the edges of the parallelogram
are given by
\begin{eqnarray}
l_{OA'}&=& l_{B'C'}= l_A= \frac{2L}{\sqrt{(4-\gamma^2)(D_1+D_2)}} \label{la} \\
l_{OC'}&=& l_{A'B'}= l_C= \frac{2L}{\sqrt{(4-\gamma^2)(D_1+D_3)}}. \label{lc} 
\end{eqnarray}
The angle $\theta$ in Fig. (\ref{fig:lt}) can be easily computed also 
\begin{equation}
\cos \theta=\frac{\gamma}{2}
\label{angle}
\end{equation}
where $\gamma$ is given in Eq. (\ref{gamma}).
Since $0\le \gamma\le 2$, it follows that $0\le \theta\le \pi/2$.

Laplace's equation (\ref{bfp1}) holds inside this parallelogram in the 
$(W_2,W_3)$ plane with the boundary conditions: $F$=1 along the edges
$OA'$ and $OC'$ and $F=0$ along the edges $A'B'$ and $B'C'$.
To find the solution, we use a conformal transformation $W(z)$ that
maps the polygon in the complex $W$ plane to an upper half complex $z$ plane.
The conformal mapping that does this is known as the Schwarz-Christoffel 
transformation. 


\section{The Schwarz-Christoffel Transformation} 
\begin{figure}
\includegraphics[height=12.0cm,width=12.0cm,angle=0]{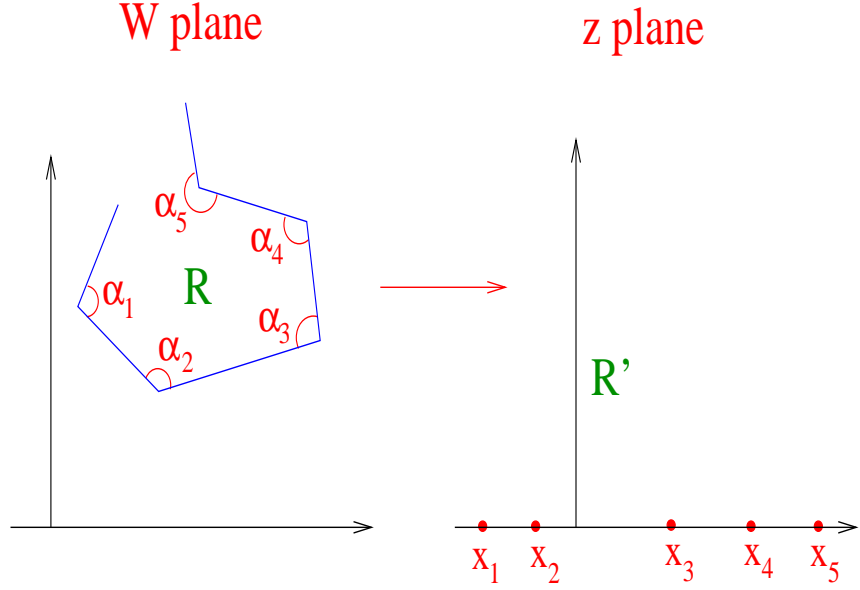}
\caption{\label{fig:sc} The Schwarz-Christoffel transformation
that maps the interior $R$ of a polygon in the complex $W$ plane
on to the upper half plane $R'$ in the complex $z$ plane.
The boundary of the polygon in the $W$ plane maps onto the real axis in the
$z$ plane.}
\end{figure}

Consider a polygon (see Fig. (\ref{fig:sc})) in the $W$ plane having 
$n$ vertices  at $\{w_1, w_2, \ldots, w_n\}$ with corresponding interior 
angles $\{\alpha_1, \alpha_2, \ldots, \alpha_n\}$. Let the points  
$\{w_1, w_2, \ldots, w_n\}$ map respectively into points 
$\{x_1,x_2,\ldots,x_n\}$ on the real axis of the $z$ plane. 
The Schwarz-Christoffel transformation $W=W(z)$ that maps the interior $R$
of the polygon in the $W$ plane on to the upper half $R'$ of the $z$ plane, 
and the boundary of the polygon on to the real axis is given by
\begin{equation}
\frac{dW}{dz}= A (z-x_1)^{\alpha_1/\pi-1}(z-x_2)^{\alpha_2/\pi-1}\ldots 
(z-x_n)^{\alpha_N/\pi-1}
\label{sc1}
\end{equation}
where $A$ is an arbitrary complex constant. Any three of the points 
$\{x_1,x_2,\ldots,x_n\}$ can be chosen at will and it is convenient to 
choose one point, say $x_n$, at infinity in which case the last factor
in Eq. (\ref{sc1}) involving $x_n$ is not present.

\begin{figure}
\includegraphics[height=12.0cm,width=12.0cm,angle=0]{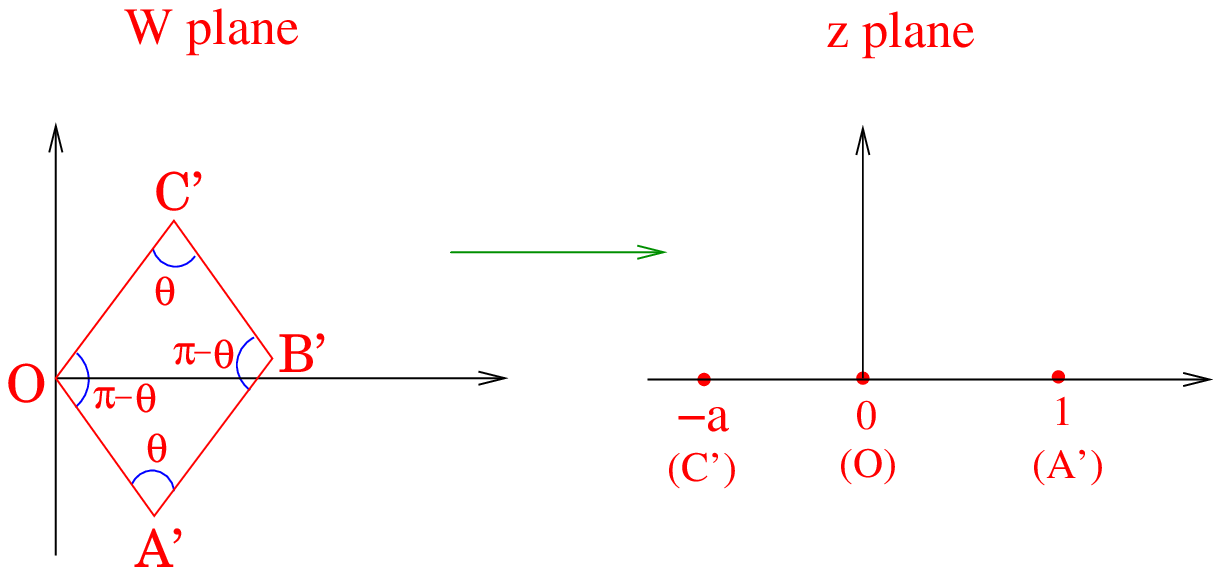}
\caption{\label{fig:ct} Under the Schwarz-Christoffel transformation
the interior $R$ of the parallelogram in the complex $W$ plane maps
on to the upper half plane in the complex $z$ plane. Under this transformation,
The boundary of the polygon in the $W$ plane maps onto the real axis in the
$z$ plane. The four vertices $C'$, $O$, $A'$ and $B'$ have their 
respective images on the real axis of the $z$ plane at $-a$, $0$,
$1$ and $\infty$.}
\end{figure}

In our problem, we have a parallelogram in the complex $W$ plane
(Fig. \ref{fig:lt}) with four vertices at $O$, $A'$, $B'$ and $C'$.
We choose three points $x_1=-a$ (image of $C'$), $x_2=0$ (image of $O$)
and $x_3=1$ (image of $A'$) and also choose the image of $B'$ to
be at infinity (see Fig. (\ref{fig:ct})). The Schwarz-Christoffel
transformation in Eq. (\ref{sc1}) then can then be written as
\begin{equation}
\frac{dW}{dz}= A (a+z)^{\theta/\pi-1} z^{-\theta/\pi} (1-z)^{\theta/\pi-1}
\label{sc2}
\end{equation}
where $A$ is still an arbitrary constant. Integrating, and using the fact
that $W(0)=0$, we get
\begin{equation}
W(z)= W_2+i W_3= A \int_0^z 
t^{-\theta/\pi}\,\left[(a+t)(1-t)\right]^{\theta/\pi-1}\, dt. 
\label{sc3}
\end{equation}

The unknown constants $A$ and the coordinate $a$ in Eq. (\ref{sc3}) are 
determined as follows.
In the complex $W=W_2+iW_3$ plane, the coordinates of the vertices
$A'$ and $C'$ are easily determined from the parallelogram
in Fig. (\ref{fig:lt}). They are respectively: 
$A'= \left(L/\sqrt{(2+\gamma)(D_1+D_2)}, -L/\sqrt{(2-\gamma)(D_1+D_2)}\right)$
and $C'=\left(L/\sqrt{(2+\gamma)(D_1+D_3)},
L/\sqrt{(2-\gamma)(D_1+D_3)}\right)$.
Under the transformation $W(z)$ they get mapped to the real $z$ axis
with coordinates $1$ and $-a$ respectively. Hence we get
\begin{eqnarray}
W(1)&=& 
\frac{L}{\sqrt{(2+\gamma)(D_1+D_2)}}-i\,\frac{L}{\sqrt{(2-\gamma)(D_1+D_2)}} 
= A \int_0^1 t^{-\theta/\pi}\,\left[(a+t)(1-t)\right]^{\theta/\pi-1}\, dt 
\label{cond1}\\
W(-a)&=& 
\frac{L}{\sqrt{(2+\gamma)(D_1+D_3)}}+i\,\frac{L}{\sqrt{(2-\gamma)(D_1+D_3)}} 
= A\int_0^{-a} t^{-\theta/\pi}\,\left[(a+t)(1-t)\right]^{\theta/\pi-1}\, dt
\label{cond2}
\end{eqnarray}
The integrals can be organized in a uniform way by defining a function
\begin{equation}
h(a) = \int_0^1 t^{-\theta/\pi}\,\left[(1+a\,t)(1-t)\right]^{\theta/\pi-1}\, dt
\label{hdef}
\end{equation}
in terms of which
\begin{eqnarray}
\frac{L}{\sqrt{(2+\gamma)(D_1+D_2)}}-i\,\frac{L}
{\sqrt{(2-\gamma)(D_1+D_2)}}&=& A 
a^{\theta/\pi-1} h(1/a) \label{t1}\\
\frac{L}{\sqrt{(2+\gamma)(D_1+D_3)}}+i\,\frac{L}
{\sqrt{(2-\gamma)(D_1+D_3)}}& =& 
-A e^{-i\theta} h(a).
\label{t2}
\end{eqnarray}
Writing $A= A_1+i A_2$ and matching the real and imaginary parts determines
$A_1$ and $A_2$ as
\begin{eqnarray}
A_1 &=& \frac{1}{h(a)} \frac{L}{\sqrt{(2+\gamma)(D_1+D_3)}} \label{A1}\\
A_2 &=& -\frac{1}{h(a)} \frac{L}{\sqrt{(2-\gamma)(D_1+D_3)}}. \label{A2}
\end{eqnarray}
This also determines $a$ via the relation
\begin{equation}
a^{\theta/\pi-1}\, \frac{h(1/a)}{h(a)}= \sqrt{\frac{D_1+D_3}{D_1+D_2}}.
\label{adet}
\end{equation}
Note that under the exchange $2\rightleftarrows 3$, $a\rightleftarrows 1/a$.

Writing $A=A_1+iA_2= |A| e^{i\beta}$, it is easy to check that
\begin{equation}
|A|= \frac{1}{h(a)} \frac{2L}{\sqrt{(4-\gamma^2)(D_1+D_3)}}
\label{amod1}
\end{equation}
which, using Eq. (\ref{adet}) can be written in a symmetrized form
\begin{equation}
|A|= \frac{2L a^{(\pi-\theta)/2\pi}}{\left[(4-\gamma^2)\,h(a)\,h(1/a)\,
\sqrt{(D_1+D_2)(D_1+D_3)}\right]^{1/2}}.
\label{amod2}
\end{equation}
The phase $\beta$ is given by, $\tan \beta= A_2/A_1= 
-\sqrt{(2+\gamma)/(2-\gamma)}$. Using $\cos \theta=\gamma/2$, one then finds
\begin{equation}
\beta= \frac{\theta-\pi}{2}.
\label{phase1}
\end{equation}
The knowledge of $A=|A|e^{i\beta}$ and $a$ (via Eq. (\ref{adet})) then fully 
determines the conformal transformation
$W(z)$ in Eq. (\ref{sc3}).

Once we have determined the appropriate conformal transformation
in \eqref{sc3}, we then need to solve Laplace's equation
$\nabla^2 F =0$ in the upper half complex $z$ plane (note that
the Laplace's equation remains invariant under conformal transformation).
The appropriate boundary conditions on the real axis of the $z$ plane
read: $F(x,0)=0$ for $x<-a$ and $x>1$ and $F(x,0)=1$ for $-a\le x \le 1$.
The solution of the Laplace's equation in the upper half $z$ plane
can be written down explicitly in terms of the boundary values by
using Poisson's formula
\begin{equation}
F(x,y)= \frac{y}{\pi}\int_{-\infty}^{\infty} 
\frac{F(x',0)dx'}{\left[y^2+(x-x')^2\right]}.
\label{Poisson1}
\end{equation}
Using our boundary conditions mentioned above and performing the integral
we get the explicit solution in the complex $z$ plane
\begin{equation}
F(z)= F(x,y)= 
\frac{1}{\pi}\left[\tan^{-1}\left(\frac{x+a}{y}\right)-
\tan^{-1}\left(\frac{x-1}{y}\right)\right].
\label{sol1}
\end{equation}
To obtain the solution in terms of the original coordinates $(y_2,y_3)$, 
we need to express $(x,y)$ in terms of $(W_2,W_3)$ 
(or equivalently $(y_2,y_3)$) using Eqs. (\ref{w2}) and (\ref{w3}), and then 
use the inverse of the conformal transformation $W(z)$ in Eq. (\ref{sc3}). 
This is rather tedious and far from illuminating. Instead in the following 
section, we derive the asymptotic solution for the distribution of the maximum
for large $L$. In this asymptotic limit, it turns out that one can explicitly 
invert the conformal transformation.

\section{Large $L$ limit: the tail of the maximum distribution}

Returning to the original cumulative distribution $F(y_2,y_3|L)$ of the 
maximum, $m$, we note that the $L$ dependence can be absorbed by rescaling 
the initial separations $y_2\to y_2/L$ and $y_3\to y_3/L$. 
In other words, the distribution
is only a function of the dimensionless variables $z_2=y_2/L$ and $z_3=y_3/L$
\begin{equation}
F(y_2,y_3|L) = F(z_2=y_2/L, z_3=y_3/L).
\label{scaling1}
\end{equation}
This means that the limit $L\to \infty$ is equivalent to taking 
limits $y_2\to 0$ and $y_3\to 0$, since $L$ always appears through 
the scaling combinations $y_2/L$ and $y_3/L$. Therefore, to extract 
the tail $L\to \infty$ of the distribution $F(y_2,y_3|L)$, we can just 
take the limits $y_2\to 0$ and $y_3\to 0$ or, equivalently, 
$W_2\to 0$ and $W_3\to 0$ in the complex $W$ plane.
This also means that we are focusing on the solution of Laplace's equation
near $z\to 0$ in the complex $z$ plane, since $W(0)=0$. The conformal 
transformation $W(z)$ in Eq. (\ref{sc3}) simplifies considerably for small $z$
since the integral for small $z$ can be trivially performed to give, 
in leading order for small $z$,
\begin{equation}
W(z) \simeq \frac{A\, a^{\theta/\pi-1}}{(1-\theta/\pi)}\, z^{1-\theta/\pi}
\label{ct:asymp1}
\end{equation} 
which, can then be easily inverted. Writing $W=W_2+iW_3=|W|^{i\psi}$, 
$z=|z|e^{i\phi}$, $A=|A| 
e^{i\beta}$ and using
$|A|$ from Eq. (\ref{amod2}), a straightforward algebra gives
\begin{eqnarray}
|z| &= &\frac{B}{L^{\pi/(\pi-\theta)}}\, |W|^{\pi/(\pi-\theta)} \label{zabs}\\
\phi &= & \frac{\pi}{\pi-\theta}(\psi-\beta) \label{phi}
\end{eqnarray}
where the constant $B$ can be expressed explicitly as
\begin{equation}
B= \sqrt{a}\,
\left(\frac{\pi-\theta}{2\pi}\right)^{\pi/(\pi-\theta)}\,
\left[
(4-\gamma^2)h(a)h(1/a)\sqrt{(D_1+D_2)(D_1+D_3)}\right]^{\pi/{2(\pi-\theta)}}.
\label{Bvalue}
\end{equation}

Once this inversion is achieved, we can take the small $z$ limit of the
explicit solution in Eq. (\ref{sol1}) that reads, to leading order,
\begin{equation}
F(x,y) \simeq 1-\frac{1}{\pi}\, \frac{1+a}{a}\, y
\label{sol2}
\end{equation}
Using $y=|z|\sin \phi$ where $|z|$ and $\phi$ are given in Eqs. (\ref{zabs})
and (\ref{phi}) respectively, we can then express the asymptotic solution as
\begin{equation}
F(y_2,y_3|L) \simeq 1- \frac{1}{\pi}\, \frac{1+a}{a}\, 
\frac{B}{L^{\pi/(\pi-\theta)}}\, |W|^{\pi/(\pi-\theta)} 
\sin\left(\frac{\pi(\psi-\beta)}{\pi-\theta}\right)
\label{sol3}
\end{equation}
where $|W|= \sqrt{W_2^2 + W_3^2}$. Using $\beta= (\theta-\pi)/2$ from 
Eq. (\ref{phase1}), one can simplify further. Finally, taking derivative 
with respect to $L$ and putting $L=m$, we obtain the tail of the 
pdf of the maximum $m$
\begin{equation}
P(m|y_2,y_3)\simeq \frac{A(y_2,y_3)}{m^{\delta}}; \quad {\rm where}\quad 
\delta= \frac{2\pi-\theta}{\pi-\theta}
\label{pdftail}
\end{equation}
and the amplitude $A(y_2,y_3)$ has the explicit expression
\begin{equation}
A(y_2,y_3)= \frac{1}{\pi} 
\left(\frac{\pi-\theta}{\pi}\right)^{\theta/(\pi-\theta)}\,
\left[\frac{(D_1D_2+D_2D_3+D_3D_1)}{\sqrt{(D_1+D_2)(D_1+D_3)}}
h(a)h(1/a)\right]^{\pi/{2(\pi-\theta)}}\,
\left(\sqrt{a}+\frac{1}{\sqrt{a}}\right)\, |W|^{\pi/(\pi-\theta)}\, 
\cos\left(\frac{\pi}{\pi-\theta}\,\psi\right)
\label{amp1}
\end{equation}
where, we recall, that $\cos \theta= \gamma/2= 
D_1/\sqrt{(D_1+D_2)(D_1+D_3)}$,
$h(a)= \int_0^1 t^{-\theta/\pi}[(1-t)(1+at)]^{\theta/\pi-1}dt$ and $a$
is determined from Eq. (\ref{adet}). In terms
of the original initial separations $y_2$ and $y_3$ we also have
\begin{equation}
|W|^2 = 
\frac{\left[(D_1+D_3)y_2^2+(D_1+D_2)y_3^2-2D_1y_2y_3\right]}
{(D_1D_2+D_2D_3+D_3D_1)}
\label{Wabs}
\end{equation}
and
\begin{equation}
\tan \psi= 
\sqrt{\frac{2+\gamma}{2-\gamma} }\,
\frac{\left[y_3\sqrt{D_1+D_2}-y_2\sqrt{D_1+D_3}
\right]}{\left[y_3\sqrt{D_1+D_2}+y_2\sqrt{D_1+D_3}\right]}\ .
\label{psi}
\end{equation}

As a check on our general result, we consider the special case when the first 
particle is immobile, i.e., $D_1=0$, and let us also assume, for simplicity,
$D_2=D_3=D$. In this case, $\gamma=0$ and hence $\theta=\pi/2$. 
The exponent $\delta= (2\pi-\theta)/(\pi-\theta)=3$.
Since,
under the exchange $D_2\rightleftarrows D_3$, $a \rightleftarrows 1/a$, it 
follows that for $D_2=D_3$, $a=1/a=1$. Hence, 
\begin{equation}
h(1)= \int_0^1 t^{-1/2}(1-t^2)^{-1/2} dt = \frac{1}{2\sqrt{2\pi}}\, 
\Gamma^2(1/4).
\label{h1}
\end{equation}
From Eq. (\ref{psi}), we have, $\tan \psi= (y_3-y_2)/(y_3+y_2)$.
Hence
\begin{equation}
\cos \left(\frac{\pi}{\pi-\theta}\psi\right)= \cos (2\psi)= 
\frac{2y_2y_3}{y_2^2+y_3^2}.
\label{psi1}
\end{equation}
From Eq. (\ref{Wabs}), we have $|W|^2= (y_2^2+y_3^2)/D$. Putting all these
expressions in Eq. (\ref{amp1}) gives,
\begin{equation}
A(y_2,y_3)= \frac{1}{4\pi^2} \Gamma^{4}(1/4)\, y_2y_3
\label{amp2}
\end{equation}
Hence, the tail of the pdf of the maximum $m$ decays as a power law
\begin{equation}
P(m|y_2,y_3)\simeq B_2 \frac{y_2y_3}{m^3}; \quad {\rm where}\quad B_2= 
\frac{1}{4\pi^2} \Gamma^{4}(1/4)=  4.37688\ldots
\label{pdftailD10}
\end{equation}
in perfect agreement with the exact result obtained for this special
case in Ref. ~\cite{KMR}. 

Let us also present the explicit result for another natural case
when all the three particles have the same diffusion constant
$D_1=D_2=D_3=D$. It follows from Eq. (\ref{bfp}) that
the distribution of $m$ is  
independent
of $D$, as $D$ drops out of the equation. In this case, 
we get from Eq. (\ref{gamma}),
$\gamma=1$ and hence $\theta=\cos^{-1}(1/2)=\pi/3$.
Hence $\delta=(2\pi-\theta)/(\pi-\theta)= 5/2$. Also, 
for $D_2=D_3$, we have $a=1$. Using this in Eq. (\ref{hdef}) and performing
the integral, we get, for $\theta=\pi/3$, $h(1)=\Gamma^2(1/3)/{2\Gamma[2/3]}$.
Then, Eqs. (\ref{pdftail}) and (\ref{amp1}) provide us the explicit
results for the tail
\begin{equation}
P(m|y_2,y_3)\simeq \frac{A(y_2,y_3)}{m^{5/2}}
\label{pdftailequal}
\end{equation}
where the amplitude is given by
\begin{equation}
A(y_2,y_3)= C 
(y_2^2+y_3^2-y_2y_3)^{3/4}\cos\left(\frac{3}{2}\tan^{-1}
\left(\sqrt{3}\frac{y_3-y_2}{y_3+y_2}\right)\right);\quad {\rm where}\quad 
C=\frac{\Gamma^{3}(1/3)}{\pi \sqrt{3} \Gamma^{3/2}(2/3)}=2.2423\dots
\label{ampequal}
\end{equation}

\section{Discussion and Summary}
In this paper we have derived the probability distribution, 
$P(m|y_2,y_3)$, for the maximum distance $m$ between the Leader and the 
Laggard, in a system of three Brownian walkers, where $y_2$ and $y_3$
are initial distances between the Leader and the other two particles. 
The probability distribution is defined over the set of processes that 
terminate when the Leader is touched (for the first time) by either of 
the other two particles. The result has, for large $m$, the power-law form
\begin{equation}
P(m|y_2,y_3) \sim A(y_2,y_3)\, m^{-\delta},
\end{equation}
where
\begin{equation}
\delta = \frac{2\pi-\theta}{\pi - \theta}
\label{delta}
\end{equation}
and $\theta$ depends on the diffusion constants via Eq.\ (\ref{theta}). 

We began this paper by discussing the seemingly unrelated problem of 
the survival probability $P(t)$, of the Leader, quoting the result 
$P(t) \sim t^{-\theta_1}$, with $\theta_1 = \pi/{2(\pi-\theta)}$, where 
$\theta$ is the same quantity that appears in Eq.\ (\ref{delta}). 
In fact we will show that the two probabilities are closely related and, 
moreover, one can determine the exponent $\delta$ by a simple scaling 
argument. 

Consider the more general function $Q(t|y_2,y_3,L)$, which is the 
{\em survival probability} of the Leader in a scenario where the process 
terminates either when the Leader is touched by one of the other two 
particles, or when one of the separations $y_2(t)$ or  $y_3(t)$ reaches 
the value $L$ (where $y_2$, $y_3$ are the initial values of these separations, 
as before). We can regard $y_2(t)$ and $y_3(t)$ as the coordinates of a 
particle diffusing inside the square $0 \le y_n(t) \le L$ ($n=2,3$). 
We define the particle as {\em surviving} if the process terminates 
by either $y_2(t)$ or $y_3(t)$ reaching the value $L$, or {\em perishing} 
if the process terminates by one of these coordinates reaching zero.  

For this general time-dependent problem, one can easily derive the backward 
Fokker-Planck equation 
\begin{equation}
\frac{\partial Q}{\partial t} = 
(D_1+D_2)\frac{\partial^2Q}{\partial y_2^2} 
+(D_1+D_3)\frac{\partial^2Q}{\partial y_3^2}
+2D_1 \frac{\partial^2Q}{\partial y_2 \partial y_3}\ ,
\label{bfpt}
\end{equation}
which is a natural generalisation of Eq.\ (\ref{bfp}). The boundary 
conditions are 
\begin{eqnarray}
Q(t|y_2=0,y_3,L)&=& 0 \label{bc10}, \\
Q(t|y_2,y_3=0,L) &=& 0 \label{bc20}, \\
Q(t|y_2=L,y_3,L) &=& 1 \label{bc30}, \\
Q(t|y_2, y_3=L,L) &= & 1 \label{bc40}.
\end{eqnarray}

Making the same change of variables as in Eqs.\ (\ref{w2}) and (\ref{w3}) 
leads to the diffusion equation
\begin{equation}
\frac{\partial Q}{\partial t} = \frac{\partial^2 Q}{\partial W_2^2} 
+ \frac{\partial^2 Q}{\partial W_3^2}\ ,
\end{equation} 
instead of the Laplace equation. In addition, the boundary conditions 
are different from (\ref{bc1}-\ref{bc4}), in that the ones and zeros 
on the right-hand side have been interchanged (due to the way we have defined 
`surviving' and `perishing'). 

After the transformation to the $W$ variables, the square domain is mapped 
to the parallelogram depicted in Figure \ref{fig:lt}. Now consider the 
the limit $L \to \infty$. In this limit the problem reduces to the calculating
the survival probability of a particle diffusing in an infinite wedge of 
opening angle $\mu = \pi -\theta$. The survival probability for 
this case is known to decay, for large $t$, as 
\cite{RednerBook,benAvraham,FisherGelfand} 
$Q(t) \sim t^{-\pi/2\mu} = t^{-\pi/(2(\pi-\theta)}$. For finite $L$, 
dimensional analysis gives, for large $t$ and $L$, 
\begin{equation}
Q_L(t) = t^{-\pi/2(\pi-\theta)} G(t/L^2)\ ,
\end{equation}
where $G(x)$ is a scaling function. In the limit $t \to \infty$, the $t$ 
dependence must drop out, giving $Q_L(\infty) \sim L^{-\pi/(\pi-\theta)}$. 
The relationship between $Q_L$ and the function $F(y_2,y_3|L)$ introduced 
in the main part of the paper is simply $Q_L = 1- F$, since both satisfy 
the same equation but with `complementary' boundary conditions (where 
the ones and zeros are interchanged between Eqs.(\ref{bc1}-\ref{bc4}) 
and Eqs.(\ref{bc10}-\ref{bc40}). We deduce that, for large $L$ 
\begin{equation}
 F(y_2,y_3|L) \to 1 - \frac{K}{L^{\pi/(\pi-\theta)}}\ ,
\end{equation}
in agreement with Eq.\ (\ref{sol3}), where $K$ is an unknown constant. 
The full solution obtained earlier fixes the value of this constant via 
Eq.\ (\ref{amp1}).  
Differentiating with respect to $L$ (and setting $L=m$) gives the probability 
distribution of the largest Leader-Laggard distance, 
$P(m|y_2,y_3) \sim m^{-\delta}$, with
$\delta = (2\pi-\theta)/(\pi-\theta)$ as in Eq.\ (\ref{pdftail}). 

We conclude by noting that the scaling analysis above as well as our exact 
solution for the three particle problem also confirms a general 
scaling result recently obtained in Ref.~\cite{MRZ} for arbitrary self-affine 
stochastic processes. Consider a self-affine
stochastic process $x(t)$ in the semi-infinite geometry ($x>0$) with
absorbing boundary condition at $x=0$. The
self-affine property simply means $x(t)\sim t^{H}$ where $H$ is
called the Hurst exponent associated with the process. Let $Q(t)$ denotes
the 
persistence probability of the process, i.e., the probability
that the process stays positive up to time $t$ and let $Q(t)\sim t^{-\theta_1}$
for large $t$, 
where $\theta_1$ is the persistence exponent~\cite{persreview}. Let $m$ denote 
the distribution
of the maximum $m$ of the process till its first-passage time through 
the origin.
Then in Ref.~\cite{MRZ}, it was argued that quite generically $P(m)\sim 
m^{-\delta}$ for large $m$ where the exponent $\delta$ is related to
the persistence exponent $\theta_1$ via the scaling relation 
\begin{equation}
\delta= 1+ \frac{\theta_1}{H}.
\label{mrzscaling}
\end{equation}
In our problem, the effective stochastic process 
$z(t)={\rm max}(y_2(t),y_3(t))$
denoting the span of the process is indeed a self-affine process with
$H=1/2$ since it represents pure diffusion. Also, from the above discussion, 
we have seen that the persistence probability $Q(t)\sim t^{-\theta_1}$ 
for large $t$ with $\theta_1=\pi/{2(\pi-\theta)}$ where $\theta$ is given 
in Eq. (\ref{theta}). Hence, the general scaling relation in 
Eq.\ (\ref{mrzscaling}) predicts that 
$\delta = 1+ 2 \theta_1= (2\pi-\theta)/(\pi-\theta)$ which
is indeed verified by the exact solution presented in this paper.
 
\begin{acknowledgments}
AB gratefully acknowledges the warm hospitality of the Laboratoire 
de Physique Th\'eorique et Mod\`eles Statistiques, Universit\'e 
Paris-Sud, Orsay, where this work was begun. 
\end{acknowledgments}

\end{document}